\documentclass[a4]{spie}
\usepackage{graphicx}

\begin{document}

\title{Strong quantum correlations in four wave mixing in $^{85}$Rb vapor}

\author{Quentin Glorieux, Luca Guidoni, Samuel Guibal, Jean-Pierre
  Likforman, Thomas Coudreau
\skiplinehalf
Universit\'e Paris-Diderot, Laboratoire Mat\'{e}riaux et
Ph\'{e}nom\`{e}nes Quantiques\\ CNRS UMR 7162\\ 10, rue A. Domon et
L. Duquet, 75013 Paris, France}

\maketitle

\begin{abstract} 
  We study quantum intensity correlations produced using four--wave
  mixing in a room--temperature rubidium vapor cell. An extensive
  study of the effect of the various parameters allows us to observe
  very large amounts of non classical correlations.
 \end{abstract}

\section{Introduction}
\label{sec:introduction}

Quantum mechanical description of the electromagnetic field introduced
45 years ago the fundamental concepts of coherent state associated to
a "standard quantum noise" that describe quantum fluctuations
associated to uncertainty principle \cite{GlauberNobel}.  At the same
time, this description clearly showed the possibility to create
squeezed states in which purely quantum correlations can reduce
quantum fluctuations associated to an observable while increasing the
fluctuations associated to its conjugate (\emph{e.g.} two quadratures
of the field) \emph{e.g.} \cite{BachorBook}.  Beyond the beautiful observation and
verification of quantum optical properties exhibited by intense beams,
the experimental realization of quantum correlated beams may become of
practical importance in order to increase the precision in optical
measurements ("noise reduction") or as a resource for quantum
information protocols.  Therefore, the search for physical systems
capable to generate large noise reduction stems from applications such
as quantum information \cite{Braunstein05} or for the observation of
very weak signals such as those expected in gravitational wave
detection\cite{Caves81}.

Four--wave mixing (4WM) in optical media displaying $\chi^{(3)}$
nonlinearities enables the generation and amplification of so--called
signal and idler beams in the presence of an intense pump beam.
Because signal and idler beams are generated by parametric conversion
of two pump photons, 4WM has been identified very early on as a very
efficient method to generate intense, non classical correlated
beams\cite{Yuen79}.  Indeed, the first demonstrations of squeezed
light were made using 4WM 25 years
ago\cite{Slusher1985,Shelby86,Maeda87}.  Soon after, it was realized
that parametric generation in $\chi^{(2)}$ crystalline media also
provided very efficient sources for the generation of intense
non--classical states\cite{Wu87,Heidmann87} with the advantages
associated to non-absorbing 
optical materials.  Indeed record amounts of quantum correlations
\cite{Laurat05} or of quantum noise reduction \cite{Mehmet10} were
produced using optical parametric amplifiers based on $\chi^{(2)}$
media.  However, recently, $\chi^{(3)}$--based experiments regained a
novel interest.  In fact, very large amounts of quantum correlations
between intense beams have been demonstrated in an experiment based on
non--degenerate 4WM in "hot" atomic vapors \cite{McCormick2007,
  McCormick08}.  These experiments have a significant advantage over
$\chi^{(2)}$ based nonclassical beam generation: the strong
nonlinearity allows for a single-pass geometry, in the absence of an
optical cavity devoted to nonlinearity enhancement in the
Continuous--Wave regime.  This is specially important for spatially
multimode quantum effects involved, for example, in quantum
imaging\cite{QuantumImaging,Boyer08}.

In this paper we present results concerning an extensive exploration
of the performances (in terms of quantum noise reduction) of
intensity--correlated beams generated by 4WM in a hot $^{85}$Rb cell
with an experimental setup largely inspired by previous successful
experiments \cite{McCormick2007, McCormick08} .  We have thus explored
in detail the parameter space, aiming to find regions where the
largest noise reductions can be found.  The paper is organized as
follows: in section \ref{sec:setup}, we give a brief description and
characterization of the experimental set--up while in
Sec.~\ref{sec:effect-parameters}, we report and comment the behavior
of the amount of quantum noise reduction as a function of the
experimental parameters, namely the temperature of the cell $T$, the
pump power, the one ($\Delta$) and two--photon ($\delta$) detunings with
respect to the atomic resonances.  This approach allowed us to obtain
up to 9.2 $\pm$ 0.5~dB of quantum noise reduction on the intensity
difference between the signal and idler modes.

\section{Experimental set--up}
\label{sec:setup}

\begin{figure}[h]
  \centering
  \includegraphics[width=.45\columnwidth]{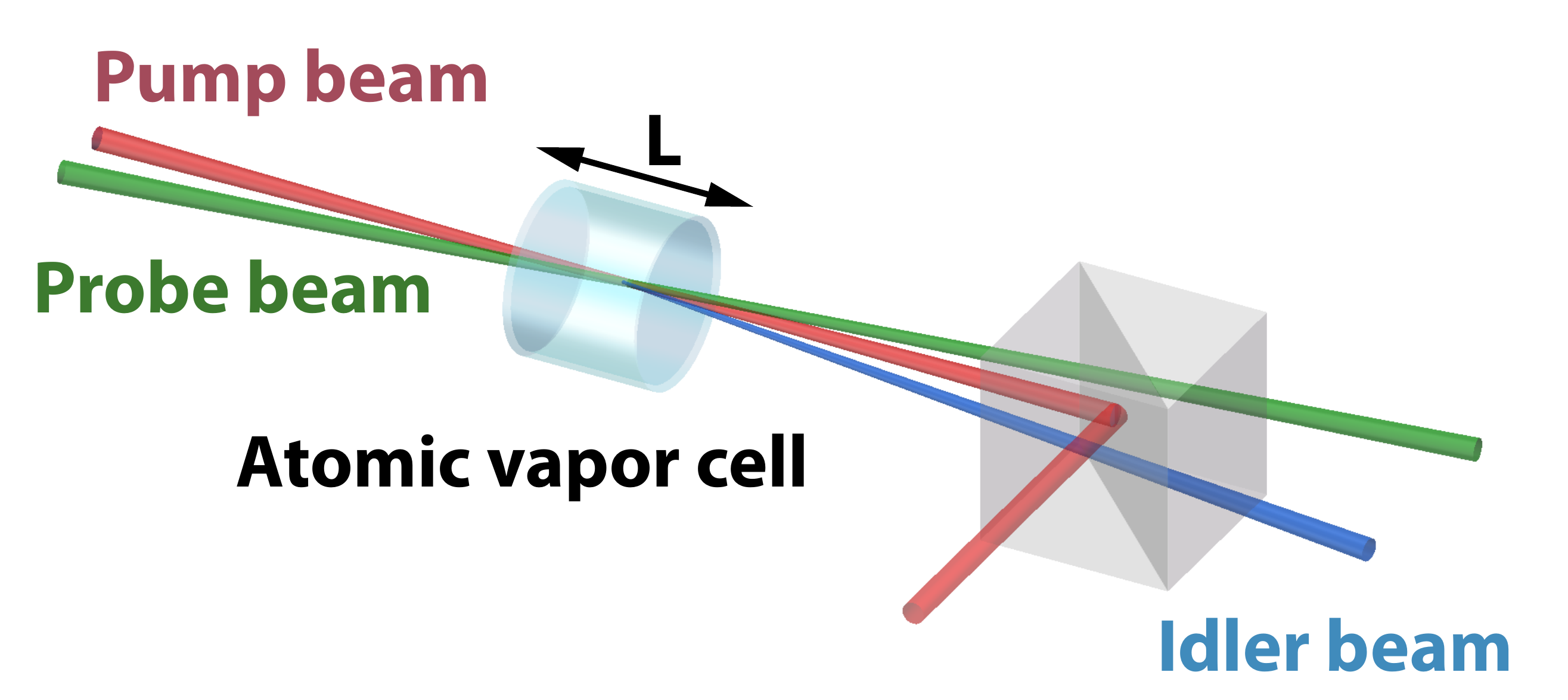}
  ~\includegraphics[width=.45\columnwidth]{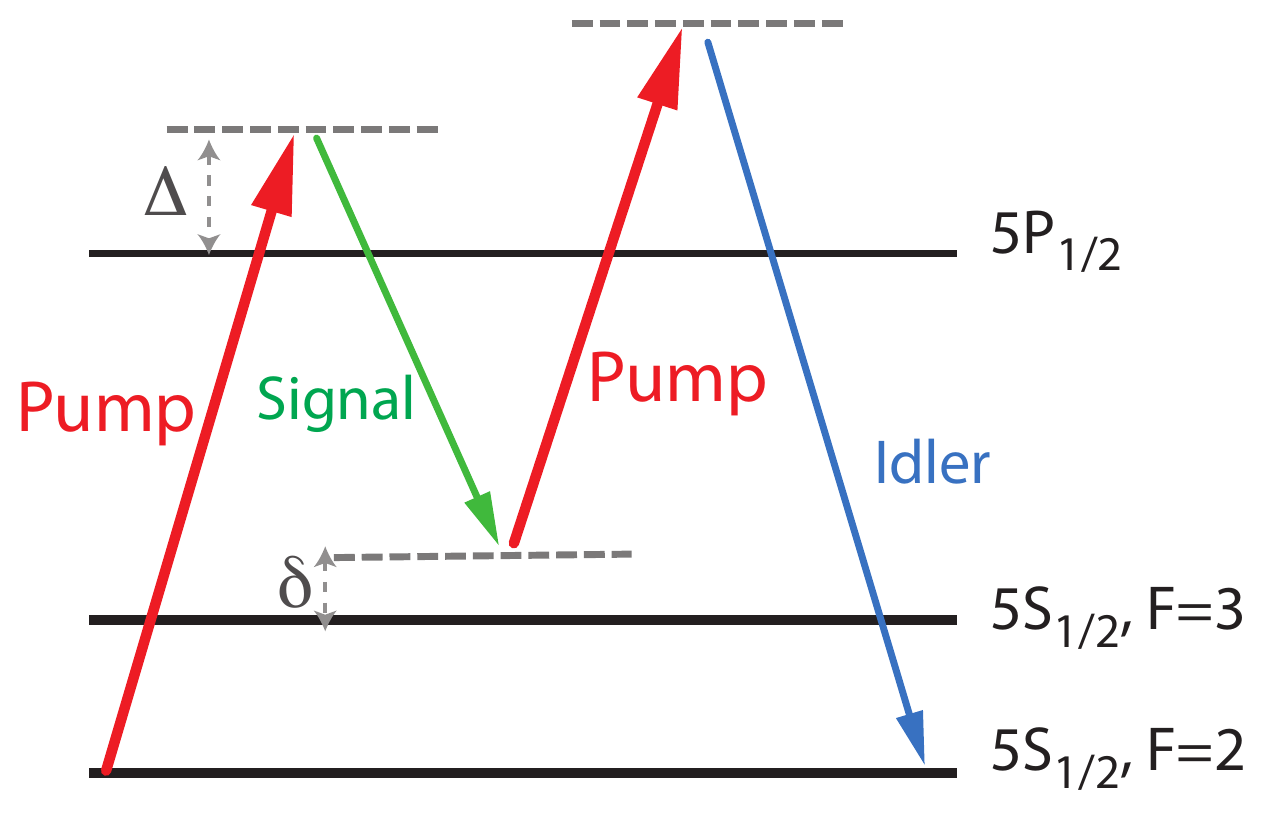} 
  \caption{Left: schematic beam path of the 4WM experiment. Right: Relevant
    levels of the $^{85}$Rb $D_1$ line. We define $\Delta$ as one--photon detuning and
    $\delta$ as two--photon detuning.}
  \label{fig:principle}
\end{figure}

The main ingredients of the experiment are schematized in
Fig.~\ref{fig:principle}: an intense pump tuned near the $D_1$ line is
mixed with a weak signal beam inside a cell containing isotopically
pure $^{85}$Rb.  As a result of the 4WM process, at the output of the
cell, the signal beam can be amplified and a conjugate beam created.
Quantum correlations between the two beams (relative--intensity
squeezing) are measured by a pair of high quantum-efficiency
photodiodes coupled to a spectrum analyzer (see
Fig.~\ref{fig:principle}).

As in the experiment presented in \cite{McCormick2007} the setup is
based on a continuous--wave ring--cavity single--mode Ti:Sa laser
(Coherent MBR, in the present case) stabilized close to the $D_1$ line
(5S$_{1/2}\rightarrow$ 5P$_{1/2}$) of $^{85}$Rb.  The laser delivers
up to 3 W with a linewidth smaller than 1~MHz.  The major part of the
power (2.5 W) can be used as a pump beam.  The probe beam is obtained
by diffracting the the remaining power in a double-pass acousto-optic
modulator (AOM) operating around 1.5 GHz (Brimrose GPF-1500) obtainig
optical powers in the range 0--300~$\mu$W.  Thus, the frequency
detuning between pump and signal beams can be adjusted around the
hyperfine splitting of the $^{85}$Rb ground state (3.036~GHz).  The
pump and probe beams are collimated over respectively 650 $\mu$m and
350 $\mu$m beam waists (radius at $1/e^2$) and crossed at very small
angle ($\simeq 1 $ mrad) inside a $L=12.5$~mm long Rubidium cell
heated at temperatures ranging from 80~$^\circ$C to
200~$^\circ$C. Pump and probe beams are linearly cross polarized so
that we can filter out most of the pump beam with a polarizing beam
splitter (extinction ratio $\simeq 10^{-5}$, Fichou).  Our detection
scheme is based on a commercial balanced detection system (Thorlab
PDB150) where two silicon photodiodes are placed in a configuration
where their photo--currents are substracted.  The difference
photocurrent is converted to tension with a transimpedance ampliflier
with a switchable gain (typically $10^5$~V/A) and sent to a spectrum
analyser (Agilent N1996A) to perform a frequency analysis at a fixed
frequency denoted $\omega$.  In the last part of this study we
replaced the original detector by a couple of two very low noise, high
quantum efficiency photodiodes that have been opened (no protecting
glass) and placed at Brewster angle.  Unless otherwise noted, all
measurements were made with a resolution bandwidth of 100~kHz and a
video bandwidth of 10~Hz.

Some particular attention has been devoted to the noise on the input signal beam.

\subsection{Input technical noise on signal beam}
\label{sec:input-noise}

\begin{figure}[h]
  \centering
  \includegraphics[width=.5\columnwidth]{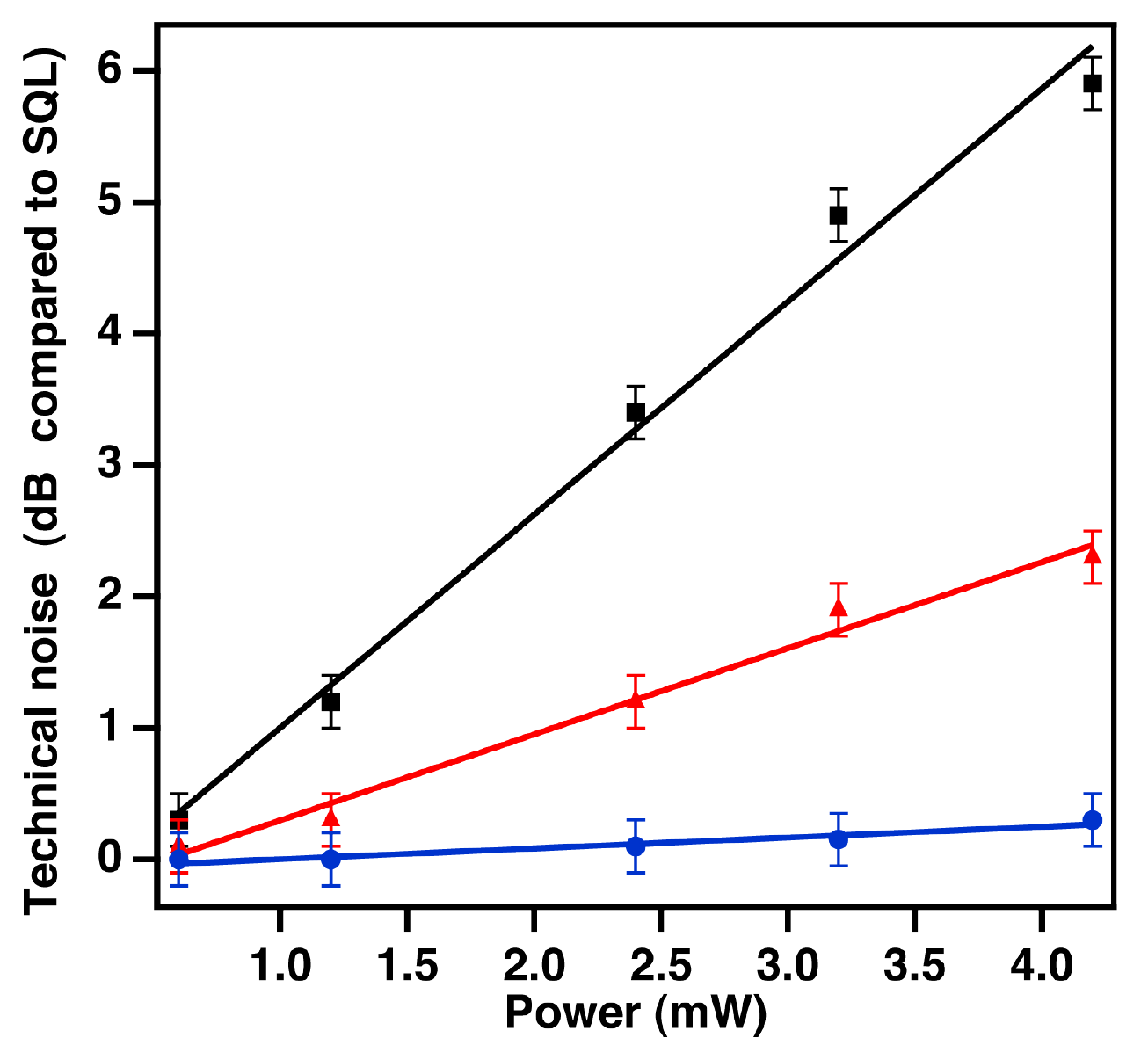}
  \caption{[Color Online]Excess noise as a function of input optical power for three
    types of RF sources, Voltage--Controlled-Oscillator (black squares),
    synthesizer and non--saturated amplifier (red triangles), synthesizer and
    saturated amplifier (blue circles).}
  \label{fig:technical-noise}
\end{figure}

Similarly to what occurs in optical parametric amplifiers, the
technical fluctuations of the input beam are important
\cite{McKenzie04}. As mentioned above, the input beam is produced by a
double pass in an AOM. As shown in Fig.~\ref{fig:technical-noise}, in
our setup the type of RF source driving the AOM plays a crucial role
on the measured fluctuations on the output beam.

Figure~\ref{fig:technical-noise} shows the excess noise on the input
signal beam at the output of the AOM as a function of the pump power
for two types of RF sources, namely
\begin{itemize}
\item an unstabilized Voltage--Controlled--Oscillator (VCO) with a
  30~dB amplifier operating in the non--saturated regime;
\item a synthesizer (Rohde \& Schwarz SMA100A) with a 30~dB amplifier operating in
  the saturated or non--saturated regime.
\end{itemize}

As can be expected, the largest amount of technical noise is observed
with the unstabilized VCO as compared to the synthesizer.  For the
synthesizer, Fig.~\ref{fig:technical-noise} shows that it is crucial
to drive the AOM with an amplifier in the saturated regime.  Let us
mention that in the case where either the VCO or the unsaturated
amplifier were used, the excess noise on the input beam prevented us
from observing noise reduction.

\section{Influence of the experimental parameters on quantum noise
  reduction}
\label{sec:effect-parameters}

Several parameters play a significant role on the observed quantum
correlations. An extensive study has been led of which we detail here
the most significant results.  Let us mention that, whereas in the
reported figures the parameters seem to span quite short intervals, we
explored a larger region in the multi-dimensional parameter--space.
We report here only the significant curves traced around the most
favorable working parameters, having in mind that for a slight change
in the parameters that are not scanned, one can easily recover quantum
noise reduction regime by re-adjusting the bias on the parameter of
interest.

\subsection{Cell temperature}
\label{sec:temperature}

The cell temperature modifies the atomic density, the Doppler width
and the transit time in the beams.

Across the presented temperature range (110$^\circ$C to 126$^\circ$C),
the Doppler width and the transit time change only by 2~\%.

On the contrary, the atomic density changes dramatically. The density
is derived from the Clausius--Clapeyron formula and assuming ideal gas
so that
\begin{equation}
  \label{eq:1}
  n_{at} = \frac{1}{k_b T} p_0 \exp( - A\times T)
\end{equation}
where $p_0$ and $A$ are given in \cite{Alcock84}. For these
parameters, the density varies between $n=1.1~10^{13}$~atoms/cm$^3$
for $T=110^\circ$~C and $n=2.8~10^{13}$~atoms/cm$^3$ for
$T=126^\circ$~C.

\begin{figure}[h]
  \centering
  \includegraphics[width=.5\columnwidth]{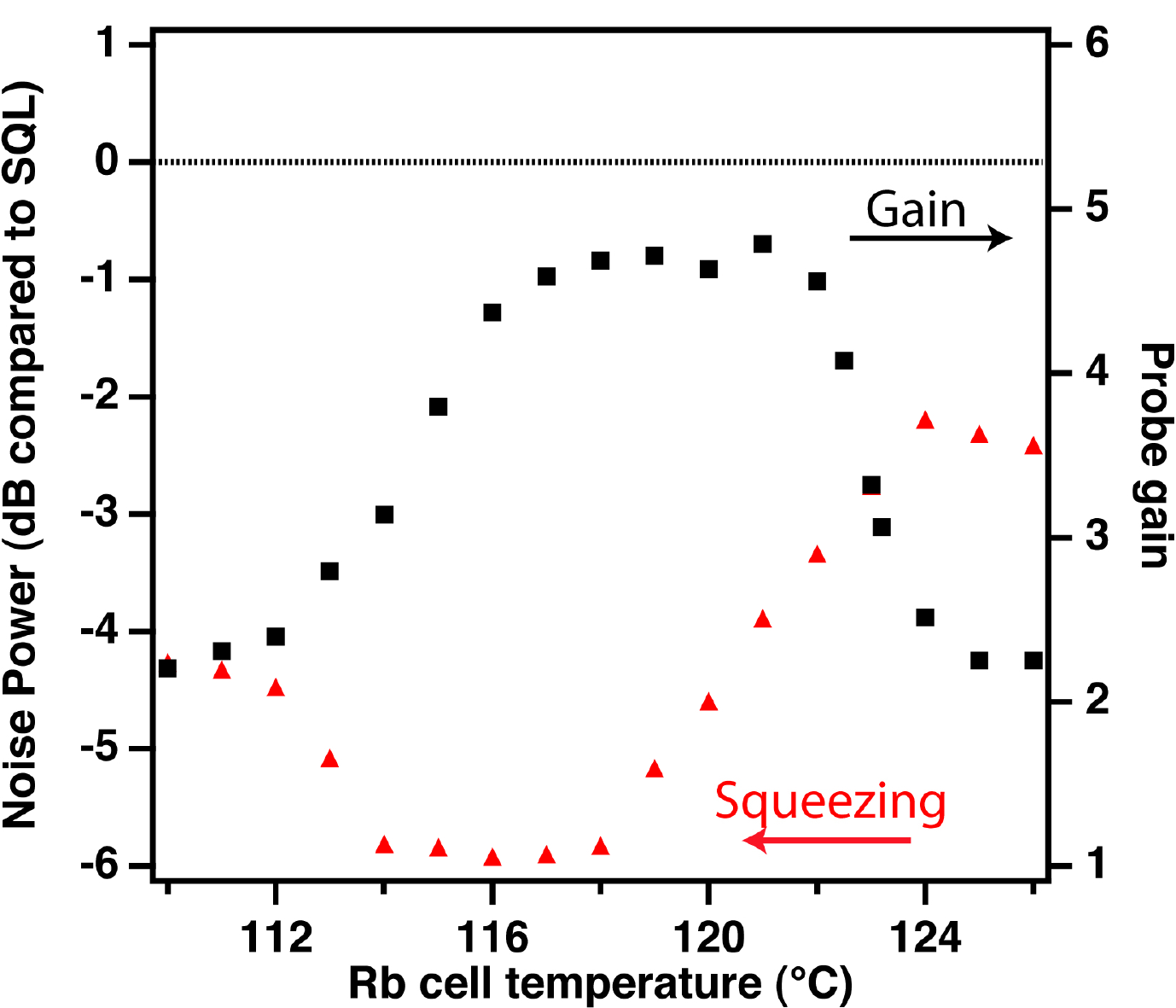}
  \caption{Normalized noise power of the intensity difference of the
    signal and idler as a function of the temperature. Other
    parameters are $P_{pump} = 500$~mW, $\delta=+10$~MHz, $\Delta =
    800$~MHz, $\omega=1.5$~MHz.}
  \label{fig:temp}
\end{figure}

The effect of the temperature and thus on the number of atoms on the
gain and quantum correlations is shown in Fig.~\ref{fig:temp}.

This curve displays an optimal region which can be understood taking
into account two effects. At low temperature / density, the number of
atoms interacting with the laser beams is too small which limits both
the gain and the quantum correlations. At higher temperature, higher
other nonlinear effects (for instance self--focusing) occur which
rapidly degrade the quantum correlations.

\subsection{One photon detuning}
\label{sec:detuning}

\begin{figure}[h]
  \centering
  \includegraphics[width=.5\columnwidth]{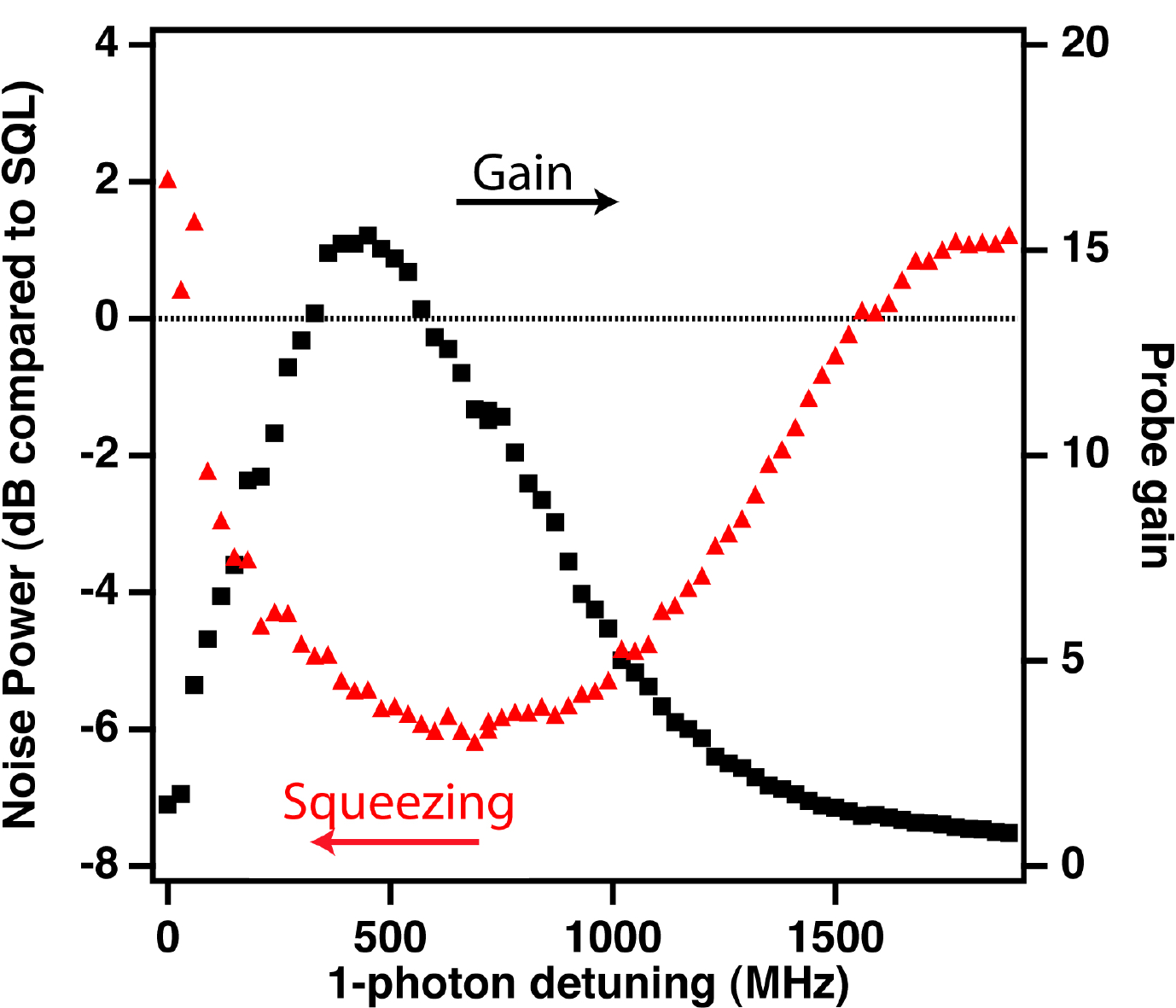}
  \caption{Normalized noise power of the intensity difference of the
    signal and idler as a function of the one--photon detuning
    $\Delta$. Other parameters are $P_{pump} = 800$~mW,
    $\delta=+10$~ MHz, $T = 114^\circ $C, $\omega=1.5$~MHz.}
  \label{fig:one-photon}
\end{figure}

Another crucial parameter for the gain and the quantum correlations is
the one photon detuning $\Delta$ between the pump and the 5S$_{1/2},
F=3 \to$ 5P$_{1/2}$ transition. Let us note that the excited level's
hyperfine structure can be neglected as it is much smaller than the
explored detuning range.  The effect of this parameter is displayed in
Fig.~\ref{fig:one-photon}.

The dependence of noise reduction on one photon detuning can be
understood in the following way. Close to resonance, absorption
becomes large which reduces the quantum correlations. Far from
resonance, the non linearity decreases thus also reducing the
correlations. The typical bandwidth is given by the Doppler broadening
(350~MHz, FWHM).

\subsection{Two photon detuning}
\label{sec:two-photon-detuning}

\begin{figure}[h]
  \centering
  \includegraphics[width=.5\columnwidth]{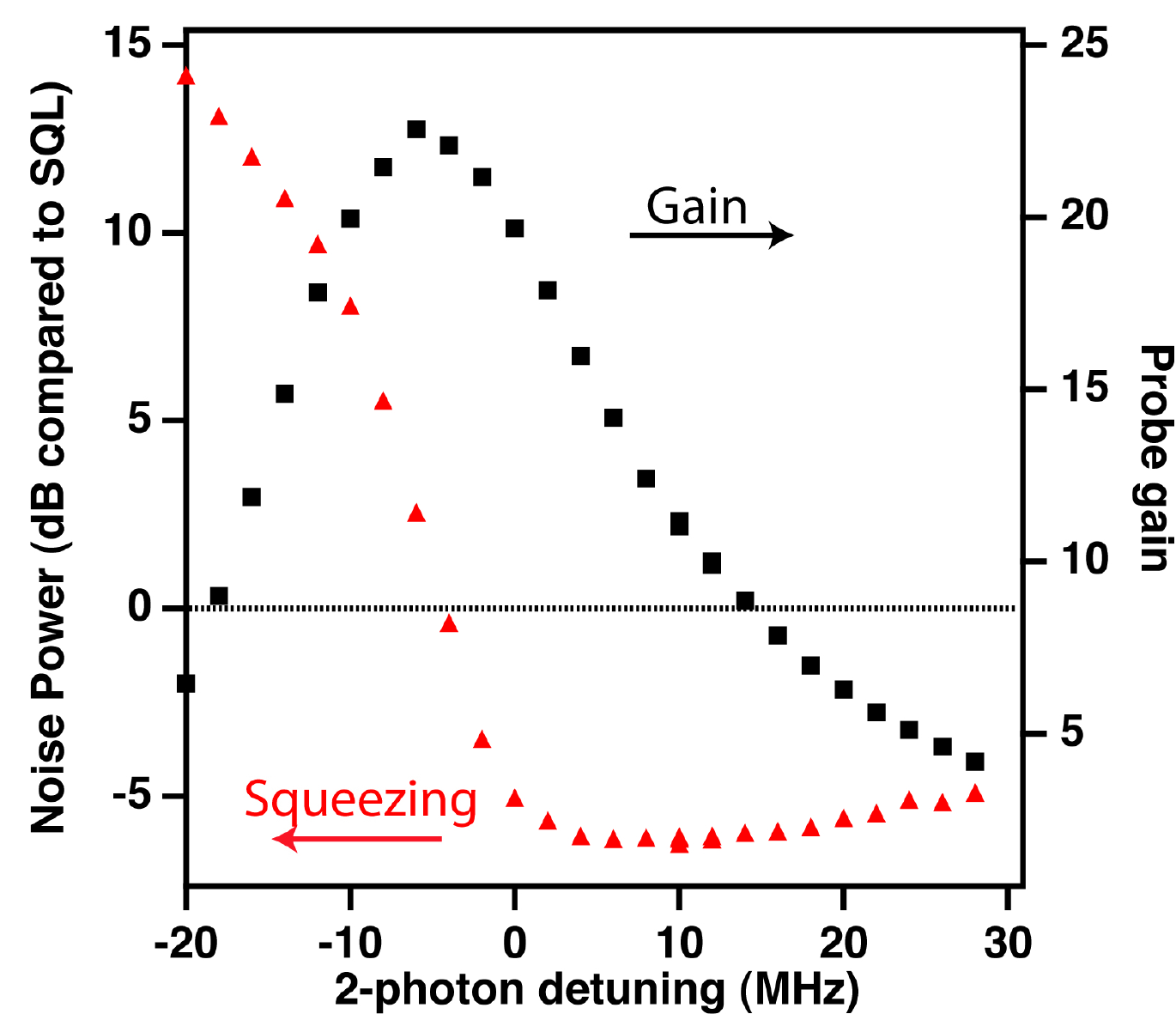}
  \caption{Normalized noise power of the intensity difference of the
    signal and idler as a function of the two--photon detuning
    $\delta$. Other parameters are $P_{pump} = 800$~mW,
    $\Delta=800$~ MHz, $T = 114^\circ $C, $\omega=1.5$~MHz}
  \label{fig:2photons}
\end{figure}

We now turn our attention to the effect of two--photon detuning,
$\delta$ (see Fig.~\ref{fig:2photons}).

Both the gain and the quantum correlations display an optimum value
around zero two--photon detuning as can be expected from resonant
enhancement. However, no simple physical picture accounts for the
observed shift ($\approx~$10~MHz) between the optimal gain and the
optimal noise reduction.

\subsection{Pump power}
\label{sec:pump-power}

\begin{figure}[h]
  \centering
  \includegraphics[width=.5\columnwidth]{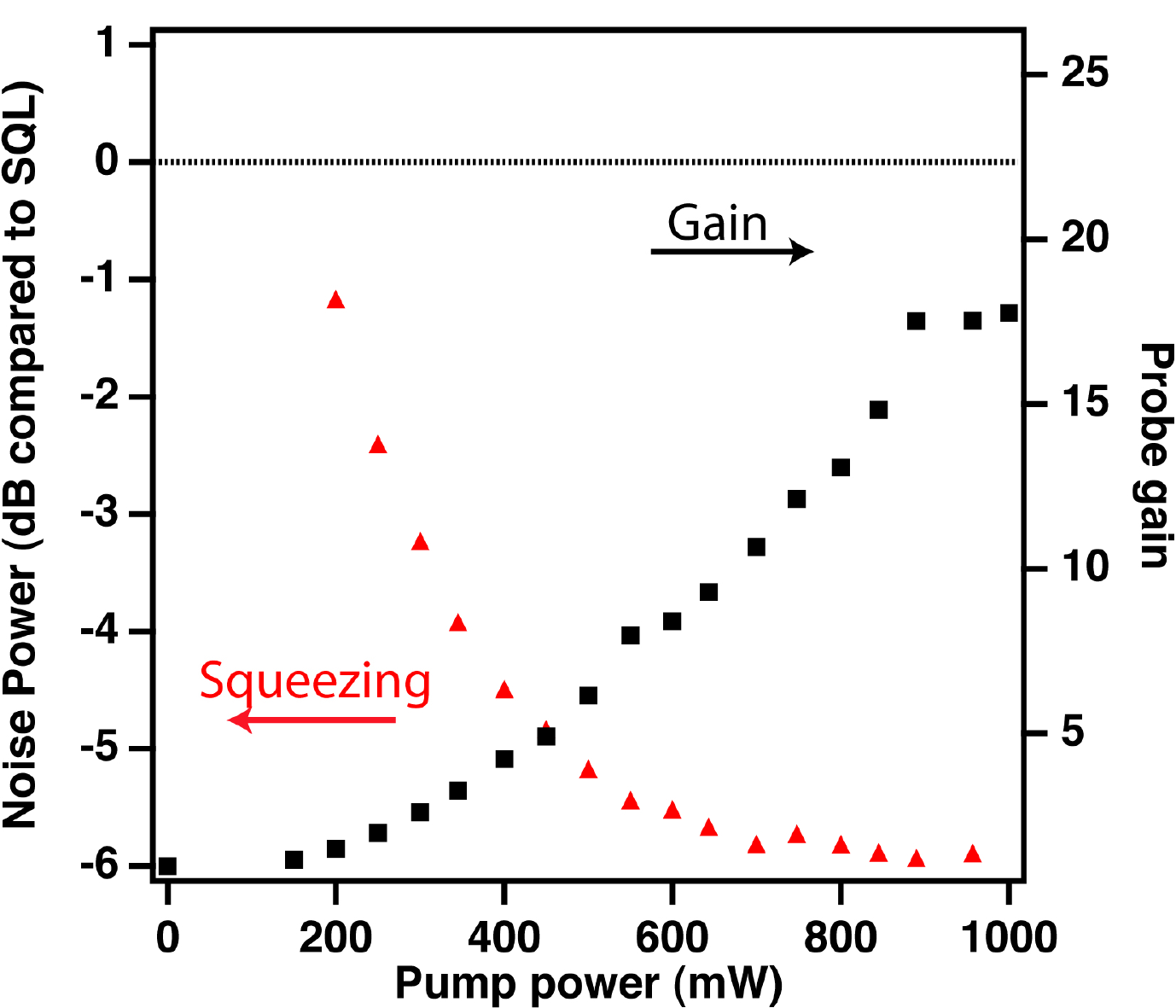}
  \caption{Normalized noise power of the intensity difference of the
    signal and idler as a function of the pump power. Other parameters
    are $\Delta=800$~ MHz, $\delta=+10$~ MHz, $T = 114^\circ $C,
    $\omega=1.5$~MHz}
  \label{fig:power}
\end{figure}

We consider now the effect of the input pump power whose effect is
displayed on Fig.~\ref{fig:power}.

In this figure, one observes that both the gain and the quantum
correlations increase with the pump power, in agreement with the
expected enhancement of the optical non linearity. Note that however
the self--focusing effects sets an experimental limit to the pump
power.

Let us note that the relevant parameter is not the absolute power but
the intensity per unit surface (proportional to the Rabi
frequency). Varying this Rabi frequency can be done either by changing
the pump power or the beam radius. However, lowering the beam radius
shortens the interaction time with the atoms which might lead to
spurious effects \cite{Lambrecht96b}.

\subsection{Optimal noise reduction}
\label{sec:optim-noise-reduct}

\begin{figure}[h]
  \centering
  \includegraphics[width=.5\columnwidth]{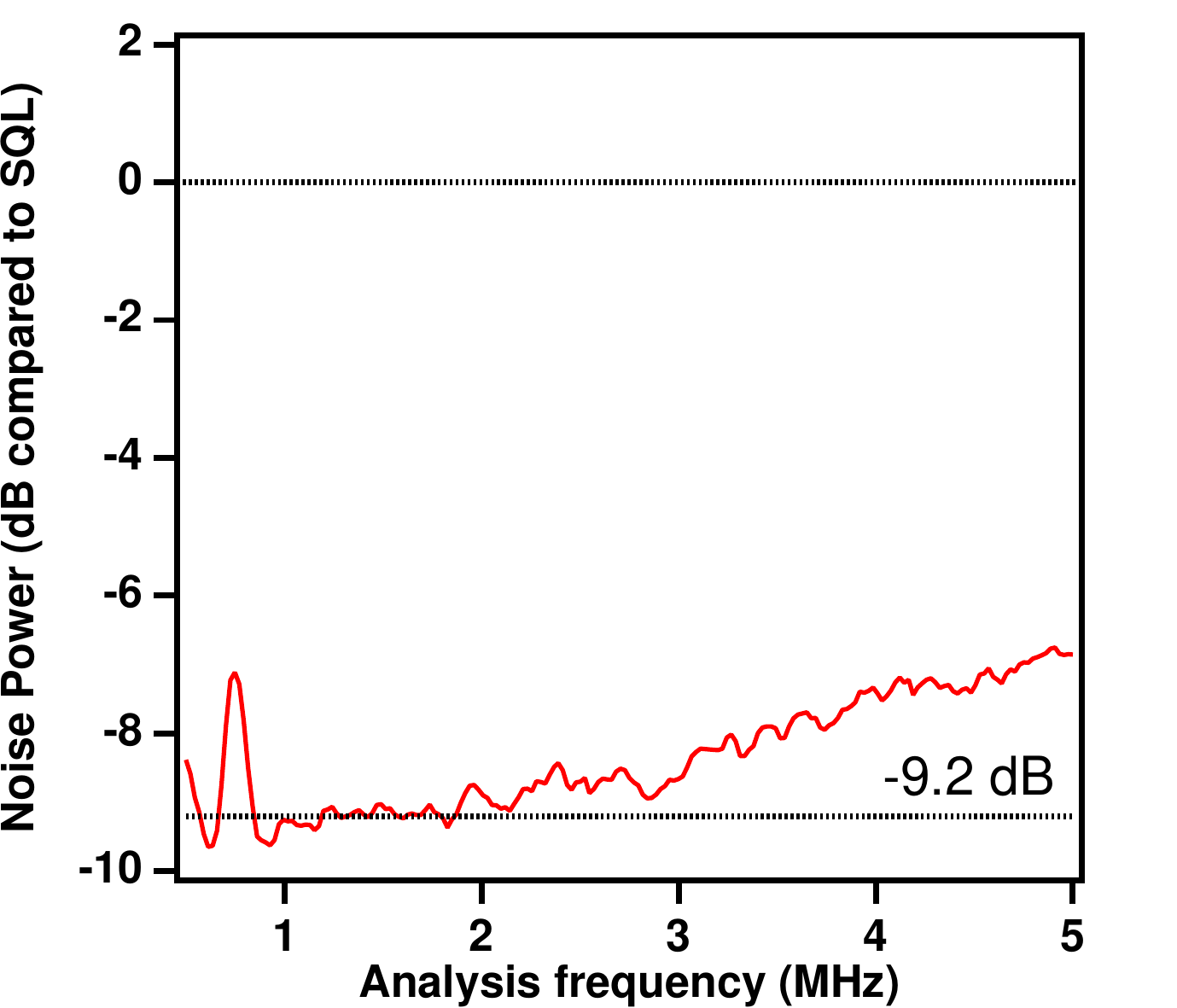}
  \caption{Noise power of the intensity difference of the signal and
    idler as a function of the frequency after correction of the
    electronic noise. A reduction of 9.2$\pm$0.5~dB is reached around 1~
    MHz. Other parameters are $\Delta=750$~ MHz, $\delta=+6$~ MHz, $T
    = 118^\circ $C, $P_{pump}=1200$~mW}
  \label{fig:record}
\end{figure}

We present here recent results obtained with improved photodiodes,
Hamatsu S3883 (in particular with the protection window removed).  We
plot in Fig.~\ref{fig:record} the noise power of the intensity
difference of the signal and idler as a function of the frequency
after correction of the electronic noise: a noise reduction of 9.2~dB
of quantum noise reduction on the intensity difference of the signal
and idler modes is observed. To our knowledge, this is the largest
noise reduction observed to date with an atomic medium.

In practice, this noise reduction is mostly limited by the losses sustained
by the two beams. These losses are the following:
\begin{itemize}
\item reflection on the cell's output window, measured at
  1.5~$\pm$~0.5~\%;
\item cumulative losses on the polarizing beamsplitter used to
  separate the pump beam from the signal and idler beams, on the iris
  used to spatially filter the signal and idler modes,  and on the
  mirrors used to deflect those two beams, estimated at 3~$\pm$~1\%;
\item limited quantum efficiency of the photodiodes. This value is
  difficult to measure as the gain of the photodetection system is unknown.
\end{itemize}
If one considers a perfect quantum efficiency for the photodiodes,
thus taking into account only the cell and propagation losses, a conservative
estimation of the noise reduction is -11.0~dB$\pm$0.7~dB below the
standard quantum limit for independent beams.

\section{Conclusion}
\label{sec:conclusion}

We have studied the effects of the temperature of the cell, the pump
power, the one and two--photon detunings with respect to the atomic
resonances on the amount of quantum correlations produced by
four--wave mixing in a rubidium vapor cell. For each of these
parameters, optimal value are presented which allow for the
observation of large quantum correlations, corresponding to
9.2~$\pm$~0.5~dB of quantum noise reduction on the intensity
difference of the signal and idler modes (-11.0~dB$\pm$0.7~dB, taking
into account the propagation losses).

Such a source provides a simple and efficient source of large CW
quantum correlations required by quantum information or quantum
measurement protocols in the absence of any optical cavity.

This study opens the way for a better comprehension of this system.
This would require a detailed model of the quantum properties of the
system including the various sources of decoherence.  Such a model
would allow for optimal conditions to be found as well as predicting
its extension to other operating conditions (other lines, alkali,
etc.)

\section*{Acknowledgments}

We thank E. Arimondo and P. Lett for fruitful discussions. 
This work was supported by Minist\`ere de l'Enseignement
Sup\' erieur et de la Recherche, Agence Nationale de la Recherche
(research contract JC05\_61454) and  Region Ile de France, project ``Communications
Quantiques''.

\end{document}